# A Higher Dimension of Consciousness

## Constructing an empirically falsifiable panpsychist model of consciousness

Version 1.0

**Jacob Jolij**

*Heymans Institute for Psychological Research,*
*Faculty of Behavioral and Social Sciences, University of Groningen,*
*The Netherlands*

j.jolij@rug.nl

## Abstract

Panpsychism is a solution to the mind-body problem that presumes that consciousness is a fundamental aspect of reality instead of a product or consequence of physical processes (i.e., brain activity). Panpsychism is an elegant solution to the mind-body problem: it effectively rids itself of the 'explanatory gap' materialist theories of consciousness suffer from. However, many theorists and experimentalists doubt panpsychism can ever be successful as a scientific theory, as it cannot be empirically verified or falsified.

In this paper, I present a panpsychist model based on the controversial idea that consciousness may be a so-called higher physical dimension. Although this notion seems outrageous, I show that the idea has surprising explanatory power, even though the model - as most models - is most likely wrong. Most importantly, though, it results in a panpsychist model that yields predictions that can be empirically verified or falsified. As such, the model's main purpose is to serve as an example how a metaphysical model of consciousness can be specified in such a way that they can be tested in a scientifically rigorous way.

# Introduction

The Mind-Body problem is arguably one of the oldest problems in philosophy. In contemporary consciousness research, it is probably best known the way Dave Chalmers (1995) phrased it, as 'The Hard Problem of Consciousness': why and how do conscious experiences exist in this universe of matter and energy? This problem has been ignored for several decades in psychology, but since the late 1990s, there has been an intensive search for the 'neural correlate of consciousness' (Crick and Koch, 1992). This renewed surge in consciousness research has led to at least 14 candidate models on how the brain produces or supports conscious sensations (Signorelli, Szczotka & Prentner, 2021).

However, most of these models rely on the classical assumption that 'mind is what the brain does', or in other words, they are materialist/physicalist models that assume that consciousness can be reduced or explained by the laws of physics. This, however, still leaves Chalmers' Hard Problem unsolved, in addition to the problem that physics itself is not causally closed yet (Ney, 2016). Materialism is therefore not universally accepted by philosophers and consciousness researchers. Although most scholars reject substance dualism, on the basis that substance dualism violates the first law of thermodynamics, there are monist alternatives to materialism. Most prominent amongst these is the notion of panpsychism, the idea that consciousness is a fundamental aspect of the universe (interpreted by some that 'everything is conscious', although one may of course debate this). According to several philosophers and neuroscientists, panpsychism is a viable alternative to materialism, and possibly even the only logical one (cf. Chalmers, 1999; Goff, 2019; Koch, 2019; also see Velmans, 2008). However, a key problem with panpsychism is not only that it is very alien to how we experience the world, but it is also extremely difficult to falsify and to integrate in a coherent metaphysical worldview that also incorporates physics.

In this paper, I will lay out a proposal to integrate a panpsychist worldview into a physical model of the universe. I do not claim this model to be correct, but rather intend to demonstrate one possibility of how one could go about in coming up with a falsifiable panpsychist model of consciousness that fits in with our understanding of physics. The resulting model does not explain consciousness as such, but rather gives a potential coherent description of how physical states are coupled to mental states (sensations) that fits with the broad panpsychist framework. The core idea is that conscious sensations are not objects, particles, forces, or events, but rather a (physical) dimension of the universe. Although this idea seems outrageous at first, I will argue that this conception of conscious states as a dimension does make sense - even if not true, the notion is an interesting 'thinking tool' to further theorizing about consciousness.

## The problem

Let us first strictly define the problem most contemporary theories of consciousness are trying to solve, namely how the brain (a physical object) produces or supports consciousness (a mental phenomenon). In the field of brain and cognitive sciences, it is now widely accepted and taught that conscious states are equivalent to brain states: "mind is what the brain does", as Minsky famously stated, or "consciousness is a brain process" (Lamme, 2006). However, what brain activity results in 'consciousness', has been a topic of debate ever since cognitive neuroscientists embraced the 'quest for consciousness' (Crick and Koch, 1992).

This 'quest for consciousness' has largely been a quest for the 'neural correlates' of sensory experience: what brain processes correlate with (reports of) sensation? With 'sensation' most researchers in the brain and cognitive sciences refer to the concept of 'qualia'. In terms of Jackson's (1986) famous parable "What Mary didn't know", consciousness researchers are looking for those brain processes that allow Mary to experience color - regardless of whether she learns something new or not when first exposed to color.

Although consciousness can mean many things, ranging from 'soul' to 'sentience' or 'intelligence', the element of conscious experience is vital in all of these, most importantly our notion of what it means to be 'human'. Sytsma et al. (2021), for example, had participants read one of two vignettes about an advanced android, which in appearance and behavior was indistinguishable from an actual human being. Critically, in one vignette they mentioned that the android did not have actual experiences, but in the other vignette the android was said to have conscious experiences. After reading the vignette, participants were asked whether the android should be awarded human rights. While most participants who read the first vignette, in which the android did not have conscious sensations, responded negatively, the majority of participants who read the second vignette, in which the android did have conscious experiences, responded positively. This shows that in our folk notion of what it means 'to be human', conscious sensations play an important role.

Therefore, it seems justifiable to focus on this most basic notion of consciousness when discussing the topic. In the remainder of this paper, I will adopt this notion: 'consciousness' is used as a synonym for 'conscious experiences' or qualia.

## Adopting a subject-neutral position

Conscious states are typically believed to be equivalent to brain states. From a phenomenological point of view, this makes sense: we experience sensations as 'our own'. There is a unique element of subjectivity and 'ownership' to conscious experience. Our conscious sensations appear to be located inside our heads (Forstmann & Burgmer, 2021), and reflect reality as projected into a 'Cartesian theatre' somewhere in the skull (Dennett & Kinsbourne, 1992). In other words: our conscious sensations are very strongly tied to our own body/brain. Consequently, thinking about consciousness and how it relates to physical processes has almost exclusively been focused on the relation between an individual's brain states and that individual's phenomenological states.

However, this feeling of ownership of experience and the idea that conscious experiences are by definition accompanied by a feeling of 'ownership', or even need a subject that experiences them, may be an illusion. There are several notable examples in psychopathology and neurology which sensations are experienced, but not integrated into the subjective self. One obvious example is that of depersonalization (DSM VI), a disorder of consciousness in which the patient reports a detachment from her or his sensations. Reality is experienced as if it is happening to someone else. Interestingly, meditation, a mental practice aimed at reducing the ego and ultimately experience reality as 'selfless', is associated with an increase in depersonalization (Castillo, 1990).

Related to this example is the practice of non-dual meditation, a form of meditation requiring extensive practice and associated with a feeling of 'pure awareness'. In a state of non-duality, the meditator does no longer experience a difference between 'self' and the world (see e.g., Laukkonen & Slagter, 2021). Non-dual states have been widely reported in the literature. We cannot make any metaphysical assumptions based on such report, of course, but so we can safely conclude that a conscious experience of 'oneness with the world' without a self, does in fact exist.

There are, however, other examples. In body integrity identity disorder (BIID), patients report that sensations from their limbs do not 'belong' to them: sensory signals from one or more extremities are not integrated into the representation of the bodily self. As a result, patients feel that a limb 'does not belong' to them - a sensation that causes such discomfort that some patients opt for amputation of the limb (Giummarra et al., 2011).

BIID and depersonalization disorder may serve as examples that sensory signals processed by the brain of an individual are not strictly necessarily integrated or attributed by a 'self'. An extremely rare case of conjoined twins, the Hogan twins seems to suggest that one of the twins can experience the sensory experiences of the other. The twins are joined at the heads and have a very unusual neural anatomy. Parts of the thalamus, the subcortical nucleus where many sensory inputs arrive and are subsequently transmitted to cortical areas, are shared between both twin sisters. Although the sisters cannot look into the same direction because of the way their skulls are fused, it appears as if one twin has access to the sensations of the other via the neural link they share, even though both sisters have distinct personalities or 'selves'. In other words, this is an example of a sensory experience being experienced by two independent minds, although the case should be interpreted with great care, given the lack of formal scientific tests of the twins (Cochrane, 2021).

Summarizing, our normal everyday experience may lead us to conclude that individual brain states are equal to particular conscious states, and necessarily tied to a 'self', or agent. This, however, might not be universally true. Of course, the notion of 'disembodied' conscious experiences sounds very alien but let us attempt to adopt a 'subject neutral' stance towards consciousness. In the 'traditional' approach towards consciousness, we make the statement

$$Q_{brain} = f(P_{brain})$$

Which should be read as conscious state Q, a point in a hypothetical 'qualia space' (Stanley, 1999), is a function of brain state P, or a transformation of state space P onto space Q. Roughly said, what we attempt in contemporary consciousness science is to find an answer to how this transformation works. Please note that although I am using the word 'function' here, I am not suggesting that there is a causal relation here (as in, the brain causes consciousness): there is an equivalence. If a given brain is in state P, this corresponds to conscious state Q, or alternatively, if we exactly know someone is experiencing state Q, the associated brain of that person is in state P.

I propose to extend this statement to

$$Q_{universe} = f(P_{universe})$$

Which should be read as state of the universe P corresponds to conscious state Q. In this, conscious state Q denotes all 'individual' conscious states or qualia experienced by all conscious entities in the universe. At first sight, this extension appears to complicate the problem we are attempting to solve. However, we cannot make assumptions about 'what it feels like' to be a self or a subject of consciousness: we necessarily have to be agnostic with respect to the type of systems that might be conscious (or experience qualia, to use a more neutral term). A theory of consciousness (or better, the function that describes the equivalence between P and Q states) needs to be neutral with respect to the medium of consciousness: brains, computers, or indeed the whole universe. Of course, it can be that only unique P to Q equivalencies exist (i.e., conscious states are only equivalent to brain states), but let us for now keep all options open.

### What then are qualia?

It is not difficult to spot the fundamental problem with an approach that attempts to describe an equivalence between physical and mental states: there is an implicit notion of dualism embedded into this question. However, as noted before, substance dualism is quite widely rejected by scientists and philosophers. If we want to come up with a physicalist model of consciousness, we need to postulate that qualia are not (part of) a separate, potentially unknowable, world, but fit in with our standard model of the universe, which describes the fundamental particles and forces, and how these interact in spacetime. This gives us only a limited number of options if we wish to incorporate qualia into this model.

For example, we could argue that qualia are mediated by some sort of special 'particles'. Jibu and Yasue (1995) propose that consciousness critically depends on the interaction between hypothetical particles they call 'corticons', fundamental particles unique to the brain, and electric fields in that brain. Obviously, any empirical evidence of existence of such corticons is lacking. Moreover, it raises the question if (and how) a new class of fundamental particles came into existence upon evolution of the mammalian cortex.

An example of an alternative account is given by Keppler (2021), who proposes that consciousness is realized in the zero-point field, the universal base electric field, and that qualia are the result of the interaction of complex electromagnetic activity in specific information processing systems with this zero point field. In other words, in this view, consciousness is more akin to a (fundamental) force, namely electromagnetism. Of course, the question remains why this force is associated with consciousness, and not the nuclear forces, for example. Additionally, the 'hard question' still remains: why are some electromagnetic interactions 'conscious', but others not?

In sum, it appears difficult to fit in qualia in the standard model of physics. However, why should we think of qualia as matter/energy? As Dennett (2018) points out, the tendency to think of qualia as a 'product' or as 'stuff' complicates our thinking about consciousness. We might be looking for something that is not there. However, this still leaves us with the question where and how to fit in consciousness or qualia in a physical model of the universe.

If qualia are not matter (i.e., particles) or a fundamental force, there appear to be only one alternative: they are an integral part of spacetime - a notion that is very close to the idea of panpsychism (cf. Frankish, 2021, who notes that panpsychism is the only logical alternative to dualism if we treat qualia as 'things' that can exist independent of a psychological subject). Spacetime as we experience it has three spatial dimensions and one temporal dimension. Contemporary physics theories such string theory, however, postulate there might be as many as 11 dimensions. Carr (2015) argues that consciousness could be conceptualized as a 'higher' dimension of spacetime. Given that it is unlikely that we can understand qualia as particles or a force, this seems to be the only alternative to give consciousness or qualia a place in a strictly physical model. Although Carr's proposals are very controversial, it might therefore be worthwhile to entertain this thought a bit further. What would we gain if we, in line with Carr, conceptualize qualia as a dimension in spacetime?

The qualia dimension

Stanley (1999) proposed the concept of a qualia space: a topological space in which all possible qualia are organized. Let us take this qualia space as a template for our qualia dimension. Where, how, and why this space came into being we cannot and need not answer for now. The qualia dimension contains all possible qualia, organized in such a way that qualia that are alike are close together: the experience of blue is closer to the experience of purple than to the experience of a high C# played on a piano; all qualia experienced by me on Tuesday, October 21st, 2021, are closer together than the qualia experienced by Emperor Augustus over 2000 years prior.

Whether qualia are organized as 'integrated' or 'unified experiences' (i.e., a single point in this qualia dimension refers to one moment of a unified experience, in which all sensory, cognitive, emotional, etc., aspects of a conscious moment are integrated), or that they are organized as individual, loose elements (i.e., what we experience as an integrated moment of conscious experience refers to several points in qualia space) is an open question. Obviously, we experience the world as an integrated whole, which would point towards the first option, but there is some psychophysical evidence that suggests that this 'wholeness' of conscious experience may be an illusion, and that different elements of a conscious experience may in fact be asynchronous 'micro-consciousnesses' (Zeki and Bartels, 1998; see also Dennett and Kinsbourne, 1998.)

In a three-dimensional universe, we assign coordinates x, y, and z to a particle for a given moment t. For a 'conscious' universe, we also need to assign a coordinate in this qualia dimension: q. It is this position on the qualia dimension that gives a particular physical state its 'feel' or 'conscious state'. This q dimension is of course not an actual spatial coordinate. Because we have an experience of the passage of time, we could at least speculate that this qualia dimension has more in common with the temporal dimension than with the spatial dimensions.

Please note that this not implies in any way that every single elementary particle 'has' consciousness or has experiences. First, just as many particles can share the same y-coordinate in a three-dimensional space, it is perfectly possible that many particles (e.g., all the particles that make up my brain, or the particles that at any moment make up the 'neural correlate of consciousness' in my brain) share the same position q on the qualia axis. In other words, one quale can be associated with many different particles. Moreover, as Stanley (1999) noted, it is conceivable that qualia space contains the quale of 'no experience'. If we wish to avoid the extreme interpretation of panpsychism (namely, that for example, also rocks and trees are conscious), we could simply state that particles that are part of a non-conscious system (whatever that may be) share the q coordinate of 'no experience'.

Even if we just take this idea as a metaphor rather than a factual description of reality, it might serve as a useful 'intuition pump' (Dennett, 2013). For example, in consciousness research there are several proposals that very explicitly make the claim that consciousness cannot be reduced to just brain activity. A notable example is Noë's Out of our heads (2010), in which Noë argues that consciousness exists as the interaction between an organism and the world around it. From a strictly neuro-materialist point of view, this seems difficult to grasp: how can consciousness exist outside a brain? However, it becomes easier to understand if we rephrase this in terms of our qualia dimension: in Noë's view, to be in a specific conscious state, not only the particles of someone's brain (or neural correlate of consciousness) need to be at a particular position q on the qualia dimension, also the particles that make up the external environment need to be at that q position.

## The uncertainty principle and Schrödinger's Cat

Although thinking about a 'qualia dimension' in terms of an intuition pump may be useful, let us further explore the possibilities this concept offers when we make an attempt to integrate it into actual physical theory. First, we should note that the concept of finding a particular particle at a given location in spacetime is not as straightforward as it seems. The Heisenberg uncertainty principle makes short work of the idea that particles have a specified location in spacetime. In contemporary physics, the Schrödinger equation describes the behavior of particles over time. Using the Born rule, we can rewrite the Schrödinger equation as a function that gives us the probability of finding a given particle at a given moment:

$$P(x, y, z, t) = |\Psi(x, y, z, t)^2|$$

Given that we have introduced a new dimension q, we should extend the Schrödinger equation (and associated probability function) accordingly:

$$P(x, y, z, q, t) = |\Psi(x, y, z, q, t)^2|$$

Interestingly, this means that for a given physical state, there is not a single quale corresponding to it, but rather a probability distribution of qualia, or, vice versa, for a given quale, there is a distribution of possible physical states. As we speculated earlier, the q dimension is most likely a temporal dimension rather than a spatial dimension. At its core, the Heisenberg uncertainty relation is about the relation between knowledge of spatial position versus temporal evolution (momentum). If we assume that the q dimension is a temporal (like) dimension, it follows that pinpointing a given spatial position of a particle reduces knowledge about its path through the qualia dimension; vice versa, if we pinpoint a particle's location or rather trajectory through the q dimension, we lose information regarding its position in the spatial dimensions.

This is obviously not how we experience the world. Here we stumble upon the infamous measurement problem in quantum mechanics: the non-classical world of quantum mechanics is one of probabilities, yet our conscious experience is one of discrete objects and events. Upon its observation, the probability function associated with a particle or system of particles collapses to a definite state. The interpretation of this collapse of the wave function is still highly debated in physics and philosophy. The Schrödinger equation is known to be the best description of reality we have: since Bell proposed his theorem in 1964 (Bell, 1964), many experiments have been done showing that quantum mechanics is a complete theory, starting with Aspect, Grangier and Roger (1980). Yet it cannot explain how and why wave function collapse occurs yet. Interestingly, the idea of a 'qualia dimension' may shed some new light on this problem.

First, there are several proposed solutions for the measurement problem. The most common solution is the so-called Copenhagen Interpretation, which some physicists have mockingly called the shut up and calculate'-approach (cf. Rosenblum & Kuttner, 2011): the Copenhagen Interpretation states that wave function collapse is real and is explicitly agnostic with respect to the further metaphysical interpretation of the phenomenon. Other alternatives include the concept of decoherence: in this interpretation, the wave function of a particle does not really collapse during a measurement, but rather becomes entangled with the wave function of a much larger measurement instrument. This much larger object forces the wave function of the measured particle into an extremely skewed probability distribution, that for all practical purposes is indistinguishable from a discrete state.

There are, however, more exotic solutions. Von Neumann (1932, see also London and Bauer, 1939), for example, proposed that conscious perception itself was responsible for the collapse of the wave function. Obviously, this 'consciousness-causes-collapse hypothesis' is controversial. If true, it would mean that consciousness is placed outside known physics. Other proposals might even sound wilder to non-physicists, such as Everett's Many Worlds model (Everett, 1957), the idea that upon every quantum measurement, the universe splits according to the possible outcomes of the measurement. A lesser-known variant on the Many Worlds model is the Many Minds account, which states that upon quantum measurement only a separate discrete consciousness percept of the world splits off, while the physical world remains in its probabilistic state (Zeh, 1970).

What does the idea of qualia as a separate dimension tell us about wave function collapse? First, we should note that this model explicitly discounts consciousness as a causal factor in wave function collapse. In the qualia dimension-model, particles move through spacetime (including the qualia dimension) according to the laws of physics, but their movement through qualia space does not affect their wave function in any way.

However, there is an additional interesting observation we can make: both the Many Minds as the Many Worlds model appear to be compatible with the idea of a q dimension. Given that, as stated earlier, the q dimension might be a temporal rather than a spatial dimension, there is an uncertainty relation between a given quale or position in the q dimension and the position of a particle in the spatial dimensions. From the perspective of a conscious observer, who can be sure to experience a particular quale Q and thus knows his position in the q dimension, this means his spatial location is probabilistic rather than deterministic. In a way, this resembles the Many Worlds model (though not exactly, of course): upon a given observation of the state of the world Q, a location on the q dimension, the observer knows that this observation must associated with a distribution of spatial coordinates, and not a fixed position - a multitude of parallel spacetime solutions or 'universes', so to say. However, this also works the other way around. Suppose we know the exact spatial location of a particle or system of particles, we can only conclude that this spatial configuration must be associated with a distribution of qualia. Fascinatingly, this is exactly what Zeh's Many Minds interpretation states.

## Moving through spacetime: what makes the clock tick?

One obvious problem with the q dimension model is that in this notion, qualia are not events or processes that unfold over time and cease to exist after the event is over. Instead, all possible experiences exist at the same time, but also do not cease to exist after we experienced them. This seems odd and, in a way, unsatisfactory: a model in which everything may happen or even does happen seems not much of a scientific model at all. However, implicitly we assume in this model that a quale, a coordinate on the q dimension, is only 'experienced' if a particle is at that given location: the 'qualia' in the q dimension are not true experiences, but rather 'proto-qualia' (cf. Russell's panprotopsychism, see Torin and Pereboom, 2019). The problem, as we have seen, is that a wave function collapse is required to give particles a discrete location in spacetime, including qualia space.

A collapse in both the Many Worlds as Many Minds interpretation, also in our framework, is an illusory event - the wave function does not actually collapse; it just appears to do so from the point of an observer. What might be problematic is that q dimension-framework lacks a clear definition of what an event actually is. In our everyday experience, which in this view is primarily defined by the movement of the particles that make up our bodies through q space, concrete events happen. Our qualia are experiences of a world of discrete objects and events. Adopting the Many Worlds or Many Minds positions above would mean that we accept that everything happens and occurs in parallel, but that we simply do not experience it that way.

If we do not wish to adopt a position in which everything happens at once, we will have to introduce a collapse mechanism in the model which gives particles a definite state in q space. However, subjective collapse positions, i.e., an interpretation of wave function collapse in which the measurement or observation of an event is instrumental in triggering a collapse, are almost anathema to the idea of a qualia dimension. As we observed earlier, the idea that qualia (or rather, proto-qualia) exist as a dimension also means they cannot affect the collapse of a wave function (the function that describes the evolution of particles in spacetime, including the q dimension) in any way: it would make the Schrödinger function self-referential. This appears to rule out the idea that any subjective aspect of a measurement or observation could play a causal role in the collapse of the wave function.

The most intuitive and elegant solution for this problem is the idea of an 'objective collapse', a wave function collapse independent of measurement. Objective collapse models are models that do not assign a causal role to observation or measurement in the collapse process: under objective collapse models, collapses may also occur without measurement. Examples are Penrose's quantum gravity induced objective reduction, or the class of Girhardi-Rimini-Weber (GRW) models. The latter class has the added advantage of having some solutions that are compatible with general relativity (see Maudlin, 2009, for an extensive discussion). The GRW-framework states that the wave function of a particle will collapse at a random moment in its lifetime. For an individual particle, this probability is very low - present parameter estimates for the decay are in the range of 10,000 years, meaning that one would have to wait 10,000 years to be sure for an individual particle's wave function to collapse. However, if such a collapse occurs, the wave functions of all particles the collapsing particle's wave function is entangled with, will also collapse.

Suppose that we extend this to let us say, a human brain. The brain is made up out of countless particles. Even though the likelihood of an individual particle's wave function to collapse is very low, because of their proximity and interactions, most particles in the brain will be entangled. This means that if only one particle's wave function collapses, the wavefunctions of the particles it is entangles with will collapse as well. Given the large number of particles, we may reasonably assume that spontaneous collapses occur regularly - even within an individual brain. Each collapse is then a discrete event in which the particles of that individual brain plus its environment get a definite state in spacetime, including the qualia dimension, and thus give reality its subjective 'feel'. These spontaneous collapses need not be limited to an individual brain, of course - if we interact with the world, the wave functions of the particles in our brain become entangled with particles in the rest of the world as well. Any spontaneous collapse in the wider system we are entangled with will result in a 'moment of consciousness' in this view.

Another, more brain-oriented solution, would be to adapt Hameroff and Penrose's Orch OR model (see Hameroff and Penrose, 2014, for an overview). Orch OR, short for orchestrated objective reduction, is a biologically inspired objective reduction model, based on Penrose's concept of quantum gravity. Orch OR postulates that microtubuli in neurons in a state of quantum superposition can become entangled with each other, to such an extent that such an entangled network of microtubuli may encompass one or more brain areas. However, if the network becomes too large, the limit for quantum gravity is exceeded and the entire systems collapses to a definite state. According to Hameroff and Penrose such an objective collapse is one 'frame' of consciousness. This would fit with the q dimension idea: upon objective collapse, particles would get a definite location in qualia space as well, and thus create a moment of conscious experience. Of course, it should be noted that the key concepts in Orch OR, that is quantum gravity and the idea that microtubuli can form entangled networks in the hot and noisy environment of the brain, remain unproven.

## Non-determinism, free will, and causation

As Dennett (2018) notes, any discussion of consciousness is incomplete without at least acknowledging the issue of free will. In the context of the present idea of a qualia dimension tied to quantum mechanics, this is relevant, as free will is - potentially - an issue in the philosophy of quantum mechanics as well (cf. Rosenblum and Kuttner, 2011; see also Hardy, 2017). The idea of a q dimension is in principle agnostic with respect to the issue of free will and determinism. Although most contemporary interpretations of quantum physics are non-deterministic, there are also determinist interpretations of quantum physics (e.g. Bohm's pilot wave model), which would also be compatible with the idea of a 'q dimension'. However, the non-determinism in quantum theory creates several interesting possibilities regarding free will, intentionality, causation, and the q dimension.

First, it is important to acknowledge that 'free will' in itself is also a sensation (cf. Haggard, 2011; Wegner, 2017). As such, conscious states associated with free will, or having made a decision, are also points in qualia space. Let us for now adopt the GRW-interpretation of the q dimension model: in the GRW-interpretation, wave functions collapse at a random moment, with a non-deterministic outcome. We have already concluded that positions in the q dimension do not have any causal relation to the wave function. In other words, the 'feeling' of willing something cannot have a causal influence on the unfolding of events as such. We might therefore conclude that the universe we live in is not only undetermined, but also that we have absolutely no influence over this non-determinism whatsoever: we are adrift on the oceans of probability - a conclusion perhaps even less desirable than strict (neuro)determinism, in which at least we have a known course.

This conclusion may be a bit too fatalistic, though. First, we need to acknowledge that, despite their inherent probabilistic nature, quantum processes are governed by probability distributions: the Born rule gives us the probability of finding a particle at a given position in spacetime. That probability can be anywhere between 0 and 1; only upon collapse it becomes 0 or 1. The past trajectory of the particle, the environment, or interactions it had, all influence these probabilities: some trajectories are more likely than others. These trajectories are described the Schrödinger equation of (a system of) particles, and in the context of the present model, also include movement through qualia space. This also means that we should give position in qualia space (i.e., conscious experience) a similar standing as position physical space when considering the evolution of a particle or system of particles.

This latter conclusion is highly relevant in the discussion of free will. Let us explore this a bit further. A conscious decision involves setting an outcome we wish to obtain in the future. For example, the decision to get a glass of water involves moving my body from my office to my kitchen, and later the experience of drinking water and quenching my thirst. Given this decision, or at least, feeling the conscious sensation to do so, this outcome is also rather likely - typically, once I have decided to go get a glass of water, I will do so. In terms of the q dimension model, this is interesting: it means that once the particles that make up my body have occupied the position of the quale 'wanting a glass water', the trajectory of these particles through spacetime is now far more likely to converge with the shortest path to my kitchen, and to the q position of the experience of drinking water. The probability functions guiding the trajectories of the particles of my body have changed as compared to before being at the coordinate of 'wanting a glass of water'.

Obviously, it is a matter of debate how the particles making up my body arrived at the q position of 'wanting a glass of water'. Homeostatic processes, prior exposure to water-related stimuli, neural activity that is not associated with consciousness - the entire history of the particles that make up my body will have influenced the probability of ending up in the part of q space that makes up the sensation of wanting a glass of water. Nonetheless, if we accept a non-deterministic account of quantum physics, the wave function collapse that will have led to the actual sensation of wanting a glass of water was an undetermined event: it could also not have occurred.

Can we now say that the conscious sensation of wanting a glass of water has caused the probability of me consuming a glass of water to increase? This depends on what one thinks about the underlying ontology and epistemology of the wave functions that make up my body. It is possible to compute the trajectory of a hypothetical particle in my brain from the point in spacetime where it is at the position in q space corresponding to 'wanting a glass of water' to the point at which it is in my kitchen. As such, the path from wanting a glass of water to getting a glass of water already exists. However, we do not know whether the particle will follow that path until we know it had a definite state in the q coordinate of 'wanting a glass of water'. So, is this 'causation'? Perhaps not in the normal sense of the word - were we agnostic with respect to the actual content of conscious experience (i.e., the specific q coordinate), we would simply see a quantum system evolve over time, without any apparent teleology or agent-induced causation. However, perceived from the perspective of an agent with free will (which would be a particle or system of particles that due to its specific configuration has access to coordinates in q space that correspond with the feeling of free will), it would feel like causation.

A final interesting, but extremely speculative possibility to consider is the idea of curvatures in q space. From general relativity, we know that spacetime is curved - an effect that we can even observe (cf. Dyson, Eddington & Davidson, 1920). This curvature affects the path particles follow through spacetime. Could q space also have a curvature, making some paths more likely than others? If so, this may be another mechanism behind the feeling we experience as 'free will': free will is a point in qualia space that is like a gravity well or attractor, curving the path through qualia space (and the other spacetime dimensions) towards it. Obviously, this is a very wild and speculative idea, but nonetheless an entertaining thought.

## Information, emergence, and entropy

Interesting and entertaining thought experiments about the curvature of a qualia space aside, several big questions about consciousness and universe are not solved by invoking a qualia dimension. For example, why is the universe organized into living, conscious units with the capacity for conscious perception? Stanley's (1999) concept of qualia space explicitly contains a 'no experience' quale - the absolute zero point for consciousness. Why would any system of particles be at a different point in qualia space?

Cleeremans and Tallon-Baudry (2021) suggest that the function of consciousness is to allow 'feeling' - a position that may sound somewhat circular. Hameroff may have worded this position slightly different: in a 2015 talk on sexual reproduction in single cellular organisms, he argued that amoebae reproduce sexually because 'it feels good' (amoebae, having microtubuli, would have conscious experiences in Hameroff's interpretation of Orch OR). Although this argument may have been made slightly in jest, the message is that qualia contain information that is not present in the physical system (cf. Jackson, 1986).

Obviously, in the q dimension idea, adding an additional dimension does add information - an additional coordinate. But there is an interesting additional observation to make: qualia themselves contain information about the other spacetime coordinates. To produce particular sensations, for example, the experience of the color blue, a specific constellation of particles is required (photons of a particular wavelength, and most likely the particles that make up the neurons associated with color perception in an individual brain). This means that once we know a particle has a q coordinate corresponding to blue, we also gain information about the other particles that make up the particle's system. Although qualia in the q dimension model are fundamental properties of spacetime, we might thus also think of them as emergent properties in a functional sense.

Varley and Hoel (2021) argues that emergent properties serve a particular goal, namely the reduction of noise at lower levels of description. There are cases in which the temporal evolution of a complex system is better predicted when looking at emergent levels (e.g., a weather pattern, such as a hurricane) as compared to the constituent levels (the air molecules that make up the hurricane) because of the noise that is present at these lower levels (cf. Brownian motion of these air molecules). In these cases, we can truly speak of an 'emergent' property with its own ontology rather than a convenient way of describing a system of particles.

There is an interesting parallel with the q dimension in this case. Even though qualia are fundamental, and not emergent properties, there are locations in qualia space that would only be accessible if a system as a whole has certain properties. For example, the feeling of 'being me' most likely requires a very specific configuration of particles. Moreover, the evolution of this system - for example, should it have decided to get a glass of water in the kitchen, as mentioned above - is easier and more accurately predicted by a single trajectory through q space than by the trajectories of all its constituent particles in spacetime. In other words, there are configurations of particles in spacetime for which the location in qualia space effectively functions as an emergent property. As we have seen in the discussion free will, occupying some definite position in qualia space may alter the probability functions of the evolution of the entire entangled system. This reasoning obviously also holds for locations in qualia space that may function as emergent properties: once a system has occupied such a position in q space, the future probabilities of the evolution of its constituent particles are affected as well.

Given that we are already freely speculating at this stage, let us take things one step further. If we accept that some states in q space result in changes in the probability functions of the system of particles occupying that space, this means that this will introduce correlations in the system, but also in its interactions with the environment. In their interpretation of the Second Law of Thermodynamics, Esposito et al. (2010) demonstrate that correlations between a system and its environment are directly related to the production of entropy. Taken together, we could argue that a system that has an 'emergent' q coordinate in its history - a system we may call a 'conscious' system - is an entropy producing system.

We may note several parallels here between this conception of conscious systems as entropy producing systems and the wider debate on the relation between information processing, entropy, and consciousness. In most mainstream contemporary theories of consciousness, information processing is seen as a vital component of consciousness, with Tononi's Integrated Information Theory (IIT) as perhaps the most explicitly information-oriented theory (see Tononi et al., 2016, for a review). Information processing, though, cannot be seen independently from thermodynamics. Faist et al. (2015) have demonstrated that any form of information processing that involves discarding information (such as an AND gate) requires work: to maintain a thermodynamical equilibrium, an information processing system releases heat into the environment, thus increasing overall entropy. This is striking, as it seems to lead to a similar conclusion as we made earlier: conscious systems (interpreted as information processing systems) must be entropy producing systems.

We can easily deduce why this is the case, and where this parallel might come from: information processing leading to conscious sensation, as it occurs in biological sensory systems, does involve a significant amount of information reduction and information loss. Processing in the retina, for example, already involves summation and integration of neural signals from individual photoreceptors to ganglion cells, during which the absolute activity levels of photoreceptors is transformed to a relative (difference) signal. The original information (the firing decrease of photoreceptor cells) is quickly lost as cis-retinal in photoreceptor cells is resynthesized and new visual input is received.

Moreover, conscious perception or even introspection is not instantaneous: estimates for the latency of conscious sensation up to 500 ms have been reported (Jolij and Lamme, 2010; Scharnowski et al., 2008). In the q dimension model, the period between registration of an external stimulus by a sense organ and the actual sensation, the particles that make up the brain move through spacetime and q space toward a particular q coordinate corresponding to, for example, seeing a blue light. This trajectory of the brain-system perceiving the blue light represents the steps necessary to reach a particular q coordinate, or, in the vocabulary of cognitive neuroscience, the processing required to perceive a given stimulus. During this transition from information registered by a sense organ to a conscious percept information is necessarily lost, as noted above. This means that entropy must be generated during processing (and thus the travel of the constituent particles of the brain-system through spacetime). A similar argument also holds for the sensation of free will - a process that may also take up 200 ms or longer from the moment a decision become inevitable (Libet, 1983; Soon et al., 2008; but see also Schultze-Kraft et al., 2016).

Obviously, there is a lot more to say about the relation between entropy, information, and consciousness in the context of the q continuum model. For example, is the observation that certain points in q space actually contain information about the configuration of particles in spacetime not counterintuitive? After all, according to the Landauer principle (see Landauer, 1961), information is physical. It must therefore be encoded in a configuration of particles. How can a coordinate on a dimension contain information then? We may propose two possible answers: first, we have already established that the q dimension is more like a temporal dimension than a spatial dimension. Given that the Second Law of Thermodynamics gives a specific direction to time, a temporal coordinate does contain information about potential configurations of particles in space: since entropy increases with time, a time coordinate further from origin is more likely to be associated with higher entropy states (i.e., states in which particles are distributed more randomly over space). As such, there is at least some information in just the temporal coordinate of a system. Given that the q dimension is most likely more like a temporal than a spatial dimension, we may allow for the q coordinate to contain information as well.

A second, more metaphysical approach to the question how q coordinates relate to information, would be to consider the idea of agency or intentionality. The concept of information is directly related to uncertainty reduction (Shannon, 1948). This raises the question: whose uncertainty? Extrapolating on Maxwell's famous demon (Maxwell, 1867), Smoluchowski (1914) recognized that the Second Law of Thermodynamics could potentially be violated by an intentional observer: an agent with knowledge about a thermodynamic system is in principle able to extract work from a heat bath. Szilard (1929) demystified the role of 'intelligent beings' in this apparent violation of the Second Law of Thermodynamics and demonstrated that any system with a sort of memory (no matter how rudimentary) could play the role of 'intentional observer', as long as the information stored in this memory would be fed back into the system. Moreover, the presence of a 'memory' requires a system consisting multiple particles. Szilard (1929) showed that such a system requires at least two particles that both are entangled with a larger system (i.e., the environment or heat bath).

How does this relate to the broader questions about information, intentionality and consciousness? Memory processes, such as described by Szilard (1929), draw information from their surroundings, and thus decrease entropy in the environment, in apparent violation of the Second Law of Thermodynamics. However, this entropy is fed back into the environment as heat when it is destroyed (cf. Landauer, 1961), restoring the balance, although this only applies to systems which interact with their environment in such a way they store information about that environment, and use that information for subsequent interactions with the environment. A rock, for example, would not be a prime example of an information processing system in this context. Although it could technically be used as an information storage device (e.g., after a rock has been heated, the decrease in temperature gives some indication of the amount of time that has passed since heating it), the system of particles that makes up the rock does not actually use this information in an intentional way.

Interestingly, memory formation has been argued to play a critical role in consciousness: Lamme (2003), for example, argues that memory formation is the neural correlate of consciousness. In other words, when information extracted from the environment is permanently stored in the physical medium of the brain, this equals to a moment of consciousness. Please note that this does not mean that all information processing in the brain is conscious - as stated above, a lot of information is processed without awareness and subsequently destroyed (thus releasing heat, according to the Landauer principle). Interestingly, within the q dimension model, storing information in memory would not contradict the Second Law of Thermodynamics: the processing leading to consciousness must produce entropy, as we have seen above.

This, however, does still not answer the question why the universe evolved to a state in which matter is organized into at least 7.7 billion (and most likely infinitely more) complex systems with access to q space coordinates that support the rich inner life we experience. On this issue we can only speculate. It might be that the observation that conscious systems produce entropy is significant is this respect: the evolution of complex systems, such as organic molecules, and living matter, requires some amount of entropy. We might hypothesize that during the early evolution of the universe the first (proto)-conscious systems started producing sufficient entropy for more complex systems to arise, analogous to how the evolution of plants on Earth provided the necessary oxygen for animals to evolve. However, such speculations, though amusing, are far out of the scope of the present paper. Let us first consider whether it is at all possible to come up with a research program to verify whether the ideas proposed here may be empirically verified at all.

## Empirical verification and falsification

We have now arrived at a point where speculations regarding the model have become rather wild. It should be obvious now to the reader that the idea of consciousness or qualia as a physical dimension of spacetime has some alluring properties, and allows for interesting digressions, associations, and philosophizing. However, linking a novel theory to earlier work or post hoc 'explaining' phenomena with a novel theory is - although a necessity in theory building - relatively easy. The proof of the pudding is coming up with constraints and in particular possibilities for empirical verification and above all, falsification.

One of the main issues when testing assumptions about the metaphysical nature of the mind-brain relationship is that we are forced - or possibly, we could argue, doomed - to look at phenomena that are counterexamples of the contemporary materialist dogma that mind and brain activity are fully equivalent. Obviously so-called paranormal phenomena, collectively referred to as 'psi', come to mind. This is both interesting and slightly worrying at the same time. Parapsychological results have been met with intense skepticism, even to such an extent that they have been labeled as 'psychology's placebo condition' in the context of the replication crisis that has been plaguing psychology for the past decade (e.g. Wagenmakers et al., 2011) - a positive finding of a parapsychological effect is according to many researchers a set of results that cannot be true and therefore must signal a problem with either the execution of an experiment of the analysis of the data. Resorting to such results to further build upon the already highly speculative theory proposed here seems risky at least.

However, it should be noted that the intense skepticism exposed by some scholars may not be wholly justified. Many critics of parapsychology, whist not necessarily agreeing with conclusion regarding the interpretation of several results, do acknowledge that the procedures and methods used by academic parapsychologists are well up to the standards in scientific practice (see e.g., French, 2018; Hyman, 1995). More importantly, the assertion that specific results 'cannot be true' is a philosophical assumption - as a matter of fact, the philosophical assumption we set out to test in the first place, namely that 'the mind' or 'consciousness' is equivalent to the known physical processes occurring in the brain. As such, I will ask the reader for some open-mindedness regarding the evaluation of parapsychological research, as this area of research appears to be a very interesting venue for potential empirical tests of the q dimension model, and, vice versa, the q model provides a fruitful model to understand psi results without having to rewrite the laws of physics.

The q dimension model makes several speculative assumptions, but based on those assumptions, we can also make predictions about qualia states, and how these relate to physical events. At its core, the q dimension model starts out as a variation on epiphenomenalism, the idea that specific mental states are equivalent to specific physical states. However, the critical addition is that because of the probabilistic nature of the Schrödinger equation, the relation between physical states and mental states in the q dimension model is probabilistic, rather than deterministic as in epiphenomenalism (at least, as most neuroscientists would understand epiphenomenalism). A second assumption is the idea that qualia space or the qualia dimension is topologically organized - a less speculative assumption, based on Stanley (1999).

Together, these two assumptions allow for an interesting observation: given the probabilistic nature of the relation between mental and physical states, sometimes our experience should be 'off' - 'off' in the sense that we experience something that is out of line with most of our experiences. Given an extreme enough deviation from normal probability, (some of) the particles that make up my brain could in very rare circumstances find themselves at a q coordinate that is normally occupied by my dog, or at a q coordinate that does not match my present temporal or spatial coordinates, for example. Such experiences would appear to the subject as 'extra sensory perception': experiencing a quale normally experienced by someone else would appear as 'telepathy'; experiencing a quale that has been displaced in space would appear as 'clairvoyance'; a displacement is time could appear as 'precognition'. It is important to note, though, that under the q model, the interpretation of such occurrences as 'perception' would be wrong: it is not 'perception' in the normal sense, i.e., the intentional gathering of information from the environment in order to adapt behavior, but we would rather characterize such events (somewhat unceremoniously) as 'freak incidents', or indeed, 'exceptional experiences'.

Although such exceptional experiences are tangent on the probabilistic nature of the wave function making up an individual, this does not mean they are completely unpredictable or random if the q model is correct. Let us consider an example like telepathy, the phenomenon that information is shared between two individuals, let us call them Alice and Bob, via a non-physical way: there is no interaction in normal spacetime (i.e., the physical dimensions of the universe) - direct or indirect - between the particles that make up Alice and Bob. In the q-model, we would interpret this as a situation in which (some) the particles that make up Alice find themselves in the q space normally taken up by Bob. The probability of such a thing happening is of course larger if Alice's position in q space is close to that of Bob. This is obviously the case if Alice is physically close to Bob, but also when Alice and Bob share the same experiences (e.g., when they are watching the same TV show), as q space is topologically organized (cf. Stanley, 1999): similar qualia are closer together in q space. This leads to a specific prediction regarding anomalous experiences: these should occur more often when individuals are close together in q space. In experimental parapsychology, this is not an unusual observation. In Ganzfeld telepathy experiments, for example, participants are often asked to meditate together prior to the experiment (see e.g., Bem and Honorton, 1994), resulting in a similar state of mind, and thus a closer position in q space.

It should be noted however, that the evidence for anomalous phenomena in laboratory experiments is extremely controversial. At the very least, it should be noted that anomalous effects are deviously difficult to replicate in laboratory settings. This is indeed to be expected under the q model: anomalous experiences are the result of probabilistic processes. The best we can hope to achieve in a laboratory setting is to manipulate probabilities, and such increase the likelihood of paranormal phenomena to occur. Given the relative rarity of anomalous phenomena, though, it is very likely that any experimental verification of anomalous effects in a laboratory setting is going to require an enormous amount of data (see also Bierman et al., 2016). A better alternative to laboratory experiments might therefore be to look at occurrence of spontaneous paranormal phenomena. According to the logic laid out above, more spontaneous phenomena should be reported when many individuals share (a set of) experiences, for example during global events such as major news events.

Of course, it is interesting to note that anomalous phenomena during major news events have been reported by Nelson (2006, but see Bancel, 2017 for a critical discussion and alternative interpretation) as part of the Global Consciousness Project, although these events concern aberrations in the behaviour of random number generators, and not the frequency of spontaneous cases. However, given the abundance of archives of spontaneous cases, it should not be too difficult to do an archive study and check this prediction of the q model, although a prospective analysis of frequency of spontaneous reported anomalous experiences would of course the preferred experiment.

However, a more powerful test of the model would be to specify one or situations that cannot be explained by the physical laws operating in the physical dimensions of the universe, but that are unique to the q dimension model. In other words, a critical confirmatory test for the model would be a reliable demonstration of a correlation between q space and physical space that cannot be explained by physical laws purely operating in the physical dimensions. One possibility that comes to mind is a further investigation of anomalies in the outputs of random number generators used in parapsychological experiments, as observed by for example Von Lucadou (reviewed in Von Lucadiou, 2011). In such experiments, participants are asked to manipulate the direction of motion of a stimulus on screen using an alleged 'secret combination' of keys of a computer keyboard. Unbeknownst to the participant, however, the motion is fully controlled by a random number generator, and not by an actual secret key combination.

In these experiments, whether the participant is successful in controlling the motion of the stimulus is purely random, as is to be expected. However, Von Lucadou and others report statistical anomalies in the output of the random number generator used in the experiment, but only for the duration that a participant was actively engaging with it, albeit indirectly (Von Lucadou, 2011; Walach et al., 2020; but see also Jolij and Bierman, 2019 for a report of anomalous correlations in a setting without a hardware RNG; and Grote, 2021 for a negative finding). Most importantly, these anomalies show up as correlation between random aspects of the participants' behaviour (such as reaction times) and the output of the random number generator. However, the exact pattern of correlations cannot be predicted beforehand.

Von Lucadou and co-workers interpret these findings in the context of a 'generalized quantum theory' (Atmanspacher, Römer & Walach, 2002). According to this theory, the correlations observed by von Lucadou are entanglement correlations between the random number generator and the participant. These entanglement correlations between macroscopic systems are possible in the GQT framework, as it proposes that the Planck constant, h, can be dropped from the Heisenberg uncertainty relation. This results in macroscopic systems being able to show quantum entanglement. This entanglement is lost, however, when one attempts to use it as a signaling medium: according to Von Lucadou, the 'no-communication theorem', a theorem from quantum information theory that forbids communication using non-locality in quantum physics, also applies in generalized quantum theory. In the case of parapsychological phenomena, a 'signal' would be a useful signal from the future, for example. According to GQT, such signals are impossible, hence the great difficulties parapsychologists are having in obtaining reliable effects in the lab: whenever a phenomenon, such as precognition, is reliably present, it may be used as a 'signal', which destroys the entanglement correlations necessary for the phenomenon to occur.

GQT, however, is rather poorly specified as to how it would work in practice. For example, it does not provide a mechanism for how and why entanglement correlations within a system come to exist, or even what comprises a 'system' in the sense of GQT. According to Von Lucadou, GQT should therefore rather be seen as a 'metaphor', describing the occurrence of odd correlations in the output of random number generators, using concepts from quantum mechanics rather than (an extension of) physical theory (Von Lucadou, personal communication, 2016).

Interestingly, the q model also predicts correlations within what Von Lucadou refers to as 'psycho-physical systems' (i.e., a system in which a human interacts - in some way - with a probabilistic device such as random number generator), but for another reason than metaphorical macroscopic quantum entanglement. In the q model, systems that move through q-space necessarily produce entropy: from sensation to sensation, information the system has picked up from the environment necessarily is lost and is dissipated into the environment. Esposito et al. (2010) have demonstrated that this can be mediated by the introduction of correlations between the system and heat sinks in the environment. These correlations, surprisingly, are negative: almost paradoxically, to maintain thermodynamic equilibrium in the long term, the introduced correlations can in some cases result in a local decrease in entropy, or, in other words, an increase in information. A similar idea is introduced by Cerf and Adami (1995) in their interpretation of quantum information theory: they introduce the concept of 'negative information', or the possibility of information to flow back from an observer or measurement system into the environment to maintain an information equilibrium in quantum systems.

Let us apply this to an information processing system moving through q space, interacting with a random number generator. At least some of the information such a system produces during its trajectory is destroyed and needs to be released back into the environment. The question is: where does this information go? Esposito et al. (2010) state that negative entropy correlations dissipate quickly in a large macroscopic system and are therefore negligible; one of the reasons we do not see information spontaneously materialize in our environment (although one might argue - very speculatively - that some spontaneous psi phenomena such as extraordinary coincidences might be just that, but we will get back later to this speculation).

Let us now focus on the rather unusual case of the Von Lucadou-style parapsychological experiments, in which unexpected correlations appear in the output of random number generators. Obviously, there is no direct physical interaction (i.e., an interaction in xyz-space) between the number generators and the research participants. However, if we look at the macrosystem, consisting of the particles making up the participant, and the particles making up the RNG, there must be an interaction in the q dimension. Given that the participant does indeed interact with the RNG, this does not seem completely unreasonable. The q model does not put any restrictions on particles that make up one 'moment of consciousness': as argued earlier, a 'moment of consciousness' is not restricted to an individual brain, but a function of all particles in a given system. So, the interaction or 'entanglement' between the RNG and the participant can be mediated via a force or direct interaction in the q dimension, rather than in normal spacetime.

It is important to realize that the random number generators typically used in these experiments produce random numbers by means of quantum mechanical processes, for example by sampling quantum tunneling in a Zener diode, or by using quantum optics, and that by design, they are in a maximally entropic state. In any system that includes such a number generator, and in which negative entropy is fed back into the environment, random number generators may function as the opposite of a heat sink, namely an 'information sink': it would be the first place where the negative entropy correlations may show up, or at least would be noticeable, when information in an information processing system is destroyed. This explains the odd patterns observed by Von Lucadou, but possibly also by Nelson and others: the correlations showing up in the output of random number generators are not a 'signal', but the negative entropy produced by an information processing system or agent, flowing back into an 'information sink' (i.e., the RNG).

However, we can (and should) go beyond merely explaining Von Lucadou's results to provide a test for the q model - as stated earlier, post hoc explanations are the easy part of theorizing. Using the q model, it is possible to make a prediction about the relative magnitude of the effect. We would need a setting in which participants interact with an RNG in two conditions that are as equal as possible, and only differ in the amount of 'unconscious' or 'preconscious' information processing, as this is the information that is lost or destroyed and will be dissipated back into the environment. The prediction that follows from the q model is that this dissipated information will show up in the RNG, and that the magnitude of this effect will be larger for the condition in which information is lost during 'unconscious' processing.

One important issue we need to note here is that information and information processing are terms that are used very loosely in (cognitive) psychology and neuroscience, and rarely (if ever) directly refer to information in the sense of Shannon-information. This makes quantitative predictions for the amount of entropy feedback as a result of information loss on the basis of, for example, the Lindauer principle, virtually impossible. However, it should be possible to at least construct an experiment in which the amount of information loss - even in the Shannon sense - is manipulated between conditions. Although we cannot make absolute quantitative predictions, we should be able to make relative predictions: if two conditions of information processing differ in the amount of information that is lost prior to a moment of consciousness, the condition with the larger amount of information loss should be associated with a larger anomalous effect in the number produced by the random number generator.

Many classical psychological paradigms that involve learning and/or deliberation would be suitable for such an experiment. The simplest experiment would be to have participants presented with a list of word or number pairs, after which they take a test on these word pairs. Critically, between the initial study phase and the test there needs to a break; one group of participants will be asked to study the words further (thus resulting in transfer to long term memory, and thus more conscious recall during the test phase), whereas the other group gets a distracting task in between, resulting in forgetting, or in other words, information loss. Obviously, one would need to incorporate a random number generator, but this can be done in several ways - from using the RNG to randomize the order of stimuli to generating numbers if number stimuli are used. Since the entanglement or correlation between the participant and the number generator is realized in q space, the interaction requires that the output of the number generator is in some way used to determine the conscious state of the participant, which can be done in many ways.

The prediction for this experiment is that, on average, the number generator will show larger anomalies for the group that is distracted between initial presentation of the stimuli and the test than for the group that is given the opportunity to study the items between initial presentation and test. Moreover, we would expect a negative correlation between the scores on the test and the RNG anomalies observed for an individual participant: a lower test score would be indicative of more forgetting, and thus of more information loss, which is associated with larger anomalies in the RNG output.

Interestingly, there may be real life analogies to the situations described above. Von Lucadou has linked his results to the controversial notion of synchronicity (Von Lucadou, Römer, and Walach, 2007), a concept coined by Carl Jung (1952) to describe 'meaningful coincidences'. According to Jung, such occurrences are examples of entanglement-like correlations mediated via an unknown medium, an idea he later fleshed out with physicist Wolfgang Pauli in their 'unus mundus' model. The idea that there are 'unseen' connections in the world is of course not new - it is commonplace in Eastern philosophy and made its way into Western thinking via Schopenhauer (1818), although it has been claimed that most philosophical, mystical, and spiritual traditions worldwide have some elements of this idea (see Huxley, 1945). Synchronicity, and the associated notion of a 'deeper connection' are obviously popular ideas in folk psychology (see for example Jaworski, 2011), since most people will have experienced the phenomenon at least once in their lives (see e.g., Wahbeh et al., 2021).

However, from a scientific point of view, 'remarkable coincidences' are not seen as particularly meaningful. Most instances of synchronicity can either be easily explained as simple causal relationships between events (i.e., there is no remarkable coincidence, as there is no coincidence in the first place), the post-hoc construction of meaning or even confabulation, or simply statistical flukes - given there are 8 billion people on this planet, something remarkable is bound to happen to someone every single day: every lottery has a 100% chance of having a winner; the chances of that winner being you is remotely small, however.

Nonetheless, like we have seen earlier with other exceptional experiences, the q model gives an interesting interpretation of synchronistic events. In Von Lucadou-style experiments, an RNG functions as an information sink, allowing for anomalous correlations to occur. However, information sinks might not be limited to RNGs. Any process or system with enough degrees of freedom (or a high enough entropy) may function as an information sink. Moreover, an odd feature of the q model is that systems can interact in a 'nonphysical' way, that is, not via xyz-space (of course, in the q model, qualia space is just as physical als xyz-space, so all interactions between particles are physical and mediated via the laws of physics, not by unknown or mystical forces) - that is, interactions via proximity in q space provide an additional valve for the dissipation of negative entropy.

Let us now speculatively give an account of synchronicity in the q model: experienced 'meaningful coincidences' are, following Von Lucadou's logic, expressions of information destroyed during 'unconscious' processing feeding back into a system. Interestingly, this is remarkably close to Jung's original interpretation of synchronicity: although most folk interpretations of remarkable coincidences are along the lines of "the universe telling us something" (cf. Jaworski, 2011), Jung's notion was that synchronicity reflects unconscious processes of some sort that manifest themselves in the external world (Jung, 1952).

Summing up, although the q model is very speculative and very poorly specified for now, it is possible to come up with a set of testable predictions - even though these predictions are at odds with our normal understanding of the physical world. That, one might argue, makes these predictions stronger rather than weaker. Moreover, these predictions can be tested not just 'in principle', but in a series of rather straightforward psychological experiments - somewhat perplexingly, we may have stumbled upon a metaphysical model that may be testable using the simplest of psychological experiments.

## Conclusion

In this paper, I have laid out a speculative model describing how conscious experience may be integrated with our physical worldview, namely by introducing consciousness as a (physical) dimension of the universe, a view compatible with (proto)panpsychism. Although the idea appears to be far-fetched at first sight, once we accept the idea of a 'consciousness space' or 'q dimension', through which particles move just as they move through normal spacetime, and which gives the universe its subjective 'feel' at a given moment in time, a surprisingly coherent worldview emerges that does not invoke any mysterianism to solve the mind-body problem.

Of course, the model presented here is wrong - all models are. Nonetheless, I at least believe to be useful. For example, the model works as an interesting metaphor or 'intuition pump' when thinking about equivalencies between mental and physical states and allows for thinking 'outside the brain' when considering ideas such as Noë's view that consciousness arises from interaction with the world and is therefore not confined to the physical brain (Dennett, 2013; Noë, 2015). However, the model itself may have its merits as well. First, apart from explaining a range of phenomena from physics to parapsychology, it yields explicit and testable predictions, which is a rather unique feature for a metaphysical model of consciousness. Whether the model is correct or not, I do believe that the line of reasoning I have adopted here is a useful one: by careful analysis and deduction, it should be possible to come up with models or proposals that are testable, even though they deal with the ineffable. Too many metaphysical approaches to consciousness are poorly constrained, and might have their appeal, but fall short where it comes to possibilities empirical verification or falsification.

I hope to have shown that is in fact is possible to come up with a non-materialist, in this case panpsychist, model that is - to some extent - falsifiable, which was the main goal of the present paper. Of course, this idea is not new. Arthur Schopenhauer already coined the term 'empirical metaphysics' when discussing anomalous phenomena he believed to be demonstrate the existence of a deeper reality, the world of Will (see Gerding, Van Dongen & Sneller, 2011); in the philosophy of physics, physicist-philosopher Abner Shimony recognized the immense value of quantum physics experiments for our fundamental and metaphysical understanding of the world (Shimony, 1978). I believe that an experimental metaphysics of consciousness is not only possible, but necessary if we want to truly solve or understand the mystery of consciousness, and that the model presented here may at least inspire others to come up with better models of their own.